\documentclass[a4paper,10pt]{article}
\usepackage{epsfig,amsfonts,url,hyperref}
\usepackage[OT2,OT1]{fontenc}
\newcommand\cyr{%
\renewcommand\rmdefault{wncyr}%
\renewcommand\sfdefault{wncyss}%
\renewcommand\encodingdefault{OT2}%
\normalfont
\selectfont}
\DeclareTextFontCommand{\textcyr}{\cyr}

\usepackage{graphicx}
\usepackage{color}

\newcommand{\link}[1]{\href{#1}{\textcircled{$\; \leftrightarrow $}}}
\newcommand{\beqn}{\begin{equation}}
\newcommand{\eeqn}{\end{equation}}
\newcommand{\beqna}{\begin{eqnarray}}
\newcommand{\beqnao}{\begin{eqnarray*}}
\newcommand{\eeqna}{\end{eqnarray}}
\newcommand{\eeqnao}{\end{eqnarray*}}
\newcommand{\ba}{\begin{array}}
\newcommand{\ea}{\end{array}}

\newcommand{\unit}{\relax\ifmmode{\rm 1\>\!\!\!I}\else{$\rm 1\!I$}\fi}
\newcounter{imageCounter}

\begin{document}
\def\q{Q}

\title{Semantic web applications with regard to math and environment}
\date{Sep 6, 2012}
\author{ Nadja Kutz\footnote{
    email: \protect\url{nad@daytar.de}}
  }

\maketitle

\begin{abstract}The following is an outline of possible strategies in using semantic web techniques and math with regard to environmental issues. The article uses concrete examples and applications and provides partially a rather basic treatment of semantic web techniques and math in order to adress a broader audience.
\end{abstract}

\section{Introduction}
This article is intented to adress a broad readership which includes on one hand a readership, which is not so familiar with the semantic web, on the other hand it should also hold something for semantic web experts. It is intended to adress mathematicians and non mathematicians, engineers, scientists and everybody else who might be interested. As an experiment this article had been developped openly at the wiki of the Azimuth project \link{http://www.azimuthproject.org/azimuth/show/Examples+of+semantic+web+applications+and+environment}. The Azimuth project \cite{AzProj} \link{http://www.azimuthproject.org/azimuth/show/HomePage} is a collaborative wiki for a science oriented audience which is interested in environmental issues. Unfortunately there exists (yet) no easy transcription of the Azimuth wiki syntax to latex or vice versa, which makes such experiments sofar quite time consuming.

The main purpose of this article is to talk about connections between disciplines and to provide some interdisciplinary applications which may be useful for environmental tasks. In the discussion of scientific and engineering applications and examples an emphasis is put on their respective mathematical structures. For that reason mathematics and the semantic web will be somewhat discussed more thouroughly.

The reader is thus kindly asked to skip over parts, which sound ``too trivial'' or ``too complicated'' for him or her. The symbol $\textcircled{$\; \leftrightarrow $}$ provides an internet link when read with a pdf reader. There are links without citation. They are intended for the readers convenience, i.e. their purpose is to give a quick link in case the reader wishes to remember or read more about some topics. Links with a citation are sources which are usually less well-known and sometimes rather only interesting for specialists.

The article is organized as follows. In section \ref{SemanMath} an introduction to the semantic web and in particular to mathematics within the semantic web is given. Subsections \ref{HTML}, \ref{RDF} and \ref{ontologies} are very basic descriptions of some important terms in the semantic web. These sections are mainly intended for readers who wish to get an introduction to the semantic web. Subsection \ref{RDFmathweb} deals with the appearance of mathematics in the semantic web and might be interesting for semantic web experts which are not so familiar with the way mathematical data is treated within the semantic web. 

In section \ref{SemanMathEnv} applications with regard to environmental issues are discussed. Subsection \ref{SemanMathEnvAppl} decribes a concrete example of how semantic techniques could support an eco engineering process. Subsection \ref{SemanMathEnvSources} deals with the special role data sources have in such a process. Section \ref{autoreason} is a little comment on (future) possibilities of using executable mathematics within the semantic web, it is mainly adressed at semantic math web experts and can be eventually skipped by most of the readership. Section \ref{RDFviz} touches upon the issue of visualizations in RDF applications and in particular in applications where visual support is used in a collaborative way. Subsection \ref{mimirix} describes a student project called {\em Mimirix} which was supervised by the author. Subsection \ref{dbpediaviz} describes some other applications and subsection \ref{deepamehta} describes some eventual future directions of the projects {\em Mimirix} and {\em Deepa Mehta}.

\section{Semantic web and math}
\label{SemanMath}

\subsection{HTML, internet and semantic web}
\label{HTML}

\vspace{2mm}
In the following some terms of the semantic web will be explained in a very basic manner.  The semantic web informed reader may skip this and jump to the middle of subsection \ref{ontologies}. 

\vspace{1mm}
There are currently quite some efforts to describe webcontent for ``machines'' (computers, robots etc.) for a brief overview, see e.g. the wikipedia entry on semantic webstack \link{http://en.wikipedia.org/wiki/Semantic_Web_Stack}. That is the data which is stored in the world wide web \link{http://en.wikipedia.org/wiki/World_Wide_Web} is only partially machine processable. This is to a great extend due to the fact that the meaning of terms in the Internet may be unclear to a machine. So in particular a term may have double meanings.  For example if one has the word ``Azimuth'' then a computer can eventually list all websites, where the term ``Azimuth'' occurs and thus a computer can in this way generate a kind of {\em meaning context} for that word (i.e. describe it in some sense), however the word ``Azimuth'' could appear in scientific applications in its mathematical meaning as an angular measurement, the word could however also refer for example to the socalled Azimuth project, Azimuth blog or forum  \cite{AzProj} \link{http://www.azimuthproject.org/azimuth/show/HomePage}. Usually only an additional information (a ``meta data'' or a ``meta description'') can clarify which meaning was meant. So, in this case, one either has to have a look-up table with the different meanings of a term and based on the generated {\em meaning context} could assign a meaning with a certain chance (in math: probability), or the term has to have a meta description which assigns a unique meaning, so that the term can be {\em identified uniquely}. Apart from the problem of double meanings, one could want to know more about special aspects of a term. One could for example want to know if a certain word is the name of a person. In this case one would need to supply the meta description ``person'' to that word. Let's investigate this further.

There are a couple of rather wellknown frameworks which deal with a machine processable description of data. The most prominent example is probably the HTML language \link{http://en.wikipedia.org/wiki/Html}, which describes ``things'' or ``objects'' which can e.g. be ``understood'' by web browsers. The HTML language was important for the development of the internet. The world wide web exists because of the HTML language. Meanwhile almost everybody knows that with the help of HTML one can for example connect data, like a text, with a certain ``internet location'' (i.e. an internet website). The location is indicated by a webadress (i.e. an URL). In other words the URL allows to {\em uniquely identify} an entity of data, like a text, as a ``data thing'' or ``data object'' in the Internet. Moreover HTML ``tells'' the browser how to link and access that data object. In particular this approach allows to mutually link data (objects), that is for example the text content (here the ``data'') of a website may be linked to the text content of another website. The linking of data is thus an important paradigm (see Linked Data on Wikipedia and Linked Data.org \link{http://linkeddata.org/}) within the development of the web. The semantic web uses a similar approach. Here individual terms instead of whole websites may get an {\em identification}. This identification can be a URL or more generally:  a {\em unique ressource identifier} (URI)  \link{http://en.wikipedia.org/wiki/Uniform_resource_identifier}. And as pointed out above with such unique identifications one may assign meta data to terms, which may a allow for a better machine processability of terms in the internet. The meta data is however usually not visible via a clickable link. Moreover ordinary web browsers make sofar not so much use of the meta information, this is however rather rapidly changing.

\subsection{the ressource description framework (RDF)}
\label{RDF}

The linking of objects is of course a rather old paradigm, which manifests itself last not but least in knitting, weaving etc. techniques of fabrics. The mathematical treatment of ``linked objects'' started with the development of {\em graph theory} \link{http://en.wikipedia.org/wiki/Graph_theory} around 1736. These old concepts are now partially translated for their use within the semantic web.

\vspace{1mm}

One concept for linking within the semantic web is is the socalled ressource description framework (RDF) \link{http://en.wikipedia.org/wiki/Resource_Description_Framework}. Socalled {\em RDF triples} \cite{RDF} \link{http://www.w3.org/TR/2004/REC-rdf-concepts-20040210/} can be thought of as representations of links as they exist in a pearl necklace, i.e. a {\em triple} can be thought of as an entity consisting of a pearl (or bead, or knot) with two threads attached to it. This entity is thus denoted as (thread 1, bead, thread 2). Graphically this looks like: $-\!\textcircled{}\!\bullet\!-.$ The threads may be tight together (``linked'') likewise the beads may be glued together, so in principle one could also think of a triple as being the entity (bead 1, thread, bead 2), graphically $\textcircled{}\!\!\!\rightarrow\!\!\!\textcircled{}.$ Note that with the arrow $\rightarrow$ (or the bead $\bullet$) it is possible to distinguish bead 1 and bead 2 (thread 1 and thread 1).  Moreover one can think of how one can connect triples to built up a ``fabric'' (a bit simplified: in mathematics the collection of links or such a ``fabric'' would be called a {\em graph}).  Mathematically a triple can thus be seen as the graph entity (node 1, arrow , node 2) or (vertex 1, directed edge, vertex 2). In other words a triple consists of three ordered ``things'', ``objects'' or ``entries'' namely thing 1, which ``is'' node 1 (within RDF thing 1 is usually called a {\em ressource} or {\em subject}), thing 2 which ``is'' the arrow (which is called in RDF a {\em property}, {\em predicate} or {\em connection}) and thing 3, which ``is'' node 2 (which is in RDF usually called an {\em object}). So a triple can be written as (ressource, property, object) or (subject, predicate, object) etc. Ressource and property of a {\em RDF triple} should ``have'' a URI (or a blank node) \cite{RDF} \link{http://www.w3.org/TR/2004/REC-rdf-concepts-20040210/}. Think of a URI as being a webadress like a URL. Hence the {\em subject} or the {\em predicate} should be represented by its adress, the {\em object} can have a URI or just be some ``literal'' like a number or a text and thus a typical {\em RDF triple} looks like (URIofSubject, URIofpredicate, URIofobject) or (URIofSubject, URIofpredicate, ``here is some text''). A collection of connected triples with URI (i.e. {\em RDF triples}) can be thus mathematically be seen as a labeled graph, where labels may however be allowed to be identical (i.e. the vertices and edges should be thought of as belonging to a ``multiset'' \cite{Blizard1992}). A collection of unconnected RDF triples is called a ``RDF graph''\cite{RDF} \link{http://www.w3.org/TR/2004/REC-rdf-concepts-20040210/}. However the author is not in overly in favor of this convention, since its to a great extend the connections, which make a graph interesting. It should also be pointed out that if one for example renders entities with the same URI as identical (i.e. the edges and vertices belong now to a set instead of a multiset) then one gets a network or ``fabric''-like graph in a canonical way. One should here also mention that the transition from multiset to set is often intrinsically given in the socalled serialization of the RDF data. A serialization \link{http://en.wikipedia.org/wiki/Serialization} describes how to arrange data like for storage or transmittance. Important serialization formats are XML \link{http://en.wikipedia.org/wiki/XML} and JSON \link{http://en.wikipedia.org/wiki/JSON}.

\subsection{RDF ontologies and possible uses}
\label{ontologies}

URI's for triple entries are usually collected (and usually also classified) in a kind of ``vocabulary'' also called {\em ontology}. An ontology is often stored at one website. There are various RDF notations, like for example N3  \link{http://en.wikipedia.org/wiki/Notation3}, Turtle  \link{http://en.wikipedia.org/wiki/Turtle_(syntax)}, RDFa  \link{http://en.wikipedia.org/wiki/RDFa}, JSON-LD  \link{http://en.wikipedia.org/wiki/JSON-LD}, RDF/JSON \link{http://docs.api.talis.com/platform-api/output-types/rdf-json} etc. which often include (at least partially) a kind of serialization. RDF triples have their own databases, called triple stores \link{http://en.wikipedia.org/wiki/Triple_store} and a prominent query language is e.g. SPARQL \link{http://en.wikipedia.org/wiki/SPARQL}. A wellknown family of languages for creating ontologies are the socalled Web Ontology Languages (OWLs) \link{http://en.wikipedia.org/wiki/Web_Ontology_Language} which are using the RDF/XML serialization and which comply to an OWL standard.  A popular open source editor for developping and processing OWLs is the {\em prot\'ege\'e OWL} editor \cite{protege} \link{http://protege.stanford.edu/overview/protege-owl.html}. In OWLs the vocabulary may be sorted within hierarchies.

A rather wellknown RDF ontology is for example at the project {\em DBpedia} \cite{dbpedia} \link{www.dbpedia.org}, where terms within wikipedia are being "identified" that is individual terms in Wikipedia are linked to a "meaning," which is stored at (identified with) an URL. Another example is the yago2 ontology at the project Yago-Naga \cite{YagoNaga} \link{http://www.mpi-inf.mpg.de/yago-naga/}. Moreover there are partial commercial projects or fully commercial projects, where data items are connected with a computer readable meaning, like to name only a few: the freebase project \cite{freebase}\link{http://www.freebase.com/}, Cyc \cite{Cyc}\link{http://en.wikipedia.org/wiki/Cyc}, Evi \cite{Evi} \link{http://en.wikipedia.org/wiki/Evi_(software)} etc.  It should be mentioned that there is a big commercial interest in having better machine accessible and interpretable data especially in connection with social data. Harvesting consumer/customers data in conjunction with general data (like about products, health and psychological statistics etc.) allows to better target and bind potential customers. The New York times article {\em How Companies Learn Your Secrets} \cite{NYTimesSecrets}\link{http://www.nytimes.com/2012/02/19/magazine/shopping-habits.html?_r=3&pagewanted=all} gives a good overview and outlook on nowadays practices. It should also be mentioned that there is of course a similar interest from various organisatorial bodies for all sorts of other applications, including applications which may not be a priori commercially oriented. So there is interest for process optimizations, organisatorial issues, data mining etc. See e.g. a video of an application from palantirtech.com \cite{palantirtech} \link{http://www.palantirtech.com/government/analysis-blog/insiderthreat}.

\subsection{the current RDF semantic web and math}
\label{RDFmathweb}

This subsection deals with the appearance of mathematics in the semantic web. A reader who ask him/herself why there should at all be much mathematics in the semantic web may jump to the next subsection.

\vspace{2mm}
There had been quite some attempts to give a semantic meaning to mathematical data.  There exist descriptions like Math-ML \cite{MathML}\link{http://en.wikipedia.org/wiki/MathML}, Open Math \cite{OpenMath}\link{http://en.wikipedia.org/wiki/OpenMath} or OMDoc \cite{OMDoc}\link{http://en.wikipedia.org/wiki/OMDoc} see also W3 math \cite{W3math} \link{http://www.w3.org/Math/}, however overall there seems to be not much math description within standartized RDF that is e.g. Math-ML doesn't seem to exist yet in (plain) RDF format. For a good overview see e.g. \cite{Lange2010} \link{http://linkeddata.future-internet.eu/images/0/04/FIA2010_Integrating_Mathematics_into_the_Web_of_Data.pdf}. However as pointed out by Massimo Marchiori \cite{Marchiori2003} \link{http://www.w3.org/People/Massimo/papers/2003/mathsemweb_mkm_03.pdf} a translation from Math-ML to RDF would at least partially be rather straightforward (see also the comments in a corresponding question on stackoverflow \cite{stackoverflow}\link{http://answers.semanticweb.com/questions/871/mathematical-expressions-in-rdf}). Christoph Lange \cite{Lange2010} \link{http://linkeddata.future-internet.eu/images/0/04/FIA2010_Integrating_Mathematics_into_the_Web_of_Data.pdf} describes some problems which could be involved in this. It should also be mentioned that there  exist some math RDF definitions in the xml schema \cite{xmlschema} \link{http://www.w3.org/TR/xmlschema11-2/}, which are used for example in the dbpedia datatypes RDF ontology, however for example the URI for integers at XML schema is mathematically not very detailled. There are also RDF decriptions of math related entities for the use in libraries see e.g. an example at iuk \cite{iuk}\link{http://www.iwi-iuk.org/material/RDF/1.1/}. For more math related RDF ontologies please see the article by Christoph Lange \cite{Lange2010} \link{http://linkeddata.future-internet.eu/images/0/04/FIA2010_Integrating_Mathematics_into_the_Web_of_Data.pdf}.

So there is a rather small appearance of math in the RDF world, on the other hand - as already indicated in the previous subsection -  there are meanwhile quite a lot of RDF descriptions of general data like names, cities etc. in the web and there is a big commercial interest in general machine processablity. The EU project LOD2 \cite{lod2}\link{http://www.lod2.eu} gives some overview also about some powerful tools for accessing the data like RDF browsers, search engines etc. (see also the below mentioned examples). The RDF linkability of mathematical content with a more general content outside of mathematics seems thus to be somewhat missing.

Apart from this there are a lot of graph theoretical questions reappearing in this context, which could be interesting for mathematicians.

\section{Semantic web, math, engineering and environment}
\label{SemanMathEnv}
\subsection{some applications of math, engineering and environment in  the semantic web}
\label{SemanMathEnvAppl}
The author of this article is rather critical, when it comes to the use of machines and computer, however it is clear that the observation and assessment of e.g. economic, environmental and social ongoings is done with the help of computers. Mathematical evaluation, mathematical modelling, data storage etc. is done with computers.

An important question is thus here in particular how to link data sets to mathematical tools. Alone taking the sum of some values is a mathematical operation, which needs a mathematical tool. RDF techniques may here be rather useful. This was also pointed out by Massimo Marchiori \cite{Marchiori2003} \link{http://www.w3.org/People/Massimo/papers/2003/mathsemweb_mkm_03.pdf}, he wrote however:

\vspace{3mm}
{\em The first factor to consider is merely technical, and has to do with the problem of vastity of domain.....There is probably the need to identify some critical subsets of mathematical knowledge that can benefit more from semantical structuring (pretty much the same kind of selection that MathML somehow had to do when facing the creation of an XML math dialect ); this seems even more important when we go to the higher layers (computability, typing, algebraic manipulation). Constructive mathematics here obviously will play a much more relevant role.

\vspace{2mm}
The second factor to consider is societal, and concerns the cost/benefit vicious circle.
}

\vspace{3mm}

So there is a danger in getting too general. And there is the danger to miss the human context. It is a very important question in how much semantic technologies really serve the human kind. Questions are here for example: does an application make for example scientific computations more transparent and accessible or is a certain application rather designed for enhancing the manipulation and control of humans in an undesired way?  These questions are of course rather explosive and manifold and so their general discussion is out of the scope of this article.  In the following we will concentrate thus on rather concrete applications which give also indications where and which mathematical entities may eventually be needed most in a societal context. Other applications are mentioned in \cite{Lange2010} \link{http://linkeddata.future-internet.eu/images/0/04/FIA2010_Integrating_Mathematics_into_the_Web_of_Data.pdf}. At the above mentioned Azimuth project \cite{AzProj} \link{http://www.azimuthproject.org/azimuth/show/HomePage} there are also discussions and works on climate models, which could be in need of real data, likewise economic simulations (see e.g. \cite{nadgames}\link{http://de.slideshare.net/nad0815/testing-new-toy-economiespolitical-structures-in-mmogs}) could be made more realistic a.s.o. The following examples shall in particular give mainly indications of what could be done. These rather user centered approaches may thus in particular be useful for the assessment of other applications. 

\vspace{2mm}
 Imagine an eco designer or an eco engineer or a group of ecodesigners/en\-gi\-neers who want to design a product, which shall be environmentally friendly. There are some eco-taxonomies (see for example the taxonomy matrix of the bundespreis ecodesign \cite{bundespreis} \link{http://www.bundespreis-ecodesign.de/de/ecodesign/matrix.html} (which can of course be encoded in an RDF graph)) within ecodesign or environmental engineering which guide this process, since there are a lot of things to consider when designing an environmentally friendly product. Lets assume the engineer follows the taxonomy and at one point thinks about which materials to use. Like he or she could be in the need for certain parts, some specific material, which doesn't deform too much during a temperature rise, which has a certain density or he or she could be in the need for data, which for example describes materialsources which are available in the vicinity of the designers geolocation, LCA aspects like carbon footprint and so on. So he or she could follow a link in an eco-taxonomy which links to "materials" and look for materials, while eventually wanting to include extra data like data of a concrete 3D model  (volume, texture, puffyness, curvature etc.) or a geolocation etc. In other words an ecoengineer would want to have an {\em easy access} to mostly technical data and methods and {\em easy tools} for customizing the data for his or her needs and especially with respect to environmental aspects. The engineer would eventually like to collaborate with other engineers etc. Note that this may include the search for certain products. Whoever tried to compare for example the specific heat capacities of different roof insulation products knows how cumbersome such a search may be.

For the example of materials the involved "mathematical content" (namely the data related to materials) is often rather simple, that is the content often consists rather just of numbers or data tables which are related to concrete measurements than to complicated functions. 3D model data of a design may however already contain Bezier splines, Christoph Lange \cite{Lange2010} \link{http://linkeddata.future-internet.eu/images/0/04/FIA2010_Integrating_Mathematics_into_the_Web_of_Data.pdf} mentions mathematical models for engines, which would make the mathematical content more complicated, climate models would be another example.

There are already commercial tools which partially accomplish the above engineering tasks like the tools by the company grantadesign \cite{grantadesign} \link{http://www.grantadesign.com}. These tools seem to be even connected to Life cycle assessment tools (LCA) and CAD-CAx tools like Autodesk \cite{autodesk} \link{http://en.wikipedia.org/wiki/Autodesk}. However there seems to be not so much variety yet for such kind of tools.

Moreover these tools are quite expensive, in particular they are way too expensive for someone in a developping country, which is of course in itself a problem for the promotion of better environmental solutions. It is also a question how good these tools can be kept up-to-date with changes in the data resources (like for concrete commercial products this may be an important point).

Thus one would like to have cheaper or even better an open source equivalent of similar tools for assessing material data, LCA data etc. Also without knowing the commercial softwares it is rather clear what kind of tool/applications are needed. Note at this place that {\bf the necessity of machine processability arises} in this example {\bf from the need to make technical data more accessible to humans}. Aspects of visualization methods in information \link{http://en.wikipedia.org/wiki/Mathematical_visualization}, mathematical \link{http://en.wikipedia.org/wiki/Mathematical_visualization} and scientific visualizations \link{http://en.wikipedia.org/wiki/Scientific_visualization} thus play quite a role in here too.

One problem however in finding open source variants of such tools is already the availability of data. There are some efforts to openly collect material data like e.g. core materials UK \cite{corematerials} \link{http://core.materials.ac.uk/} but these efforts are still rather small. So one has in particular to ask wether it is eventually senseful to further crowdsource the collection of the respective data and to ask how to process and use it. It is here in particular to be asked in which sense departments of materials science and engineering or organisatorial bodies as e.g. the ISO \link{http://www.iso.org} could be involved in providing data and wether existing (materials) data repositories could be involved in this.

Another data source which is also concerned with materials, however more on a physical/chemistry level is for example a materials genome properties repository from high-throughput ab-initio calculation \link{http://en.wikipedia.org/wiki/Ab_initio_quantum_chemistry_methods} like \cite{aflowlib} \link{http://www.aflowlib.org/}. Solid state materials like for example for photovoltaic cells are still to quite some extend explored via experiments rather than via computational methods. The enhancement of the computational side could eventually spur the process of finding more environmentally friendly/efficient materials. The data could here of course be supplemented with rather complex mathematical content. Note that the transparency and direct accessability of the computations themselves is here rather important.

\subsection{semantic web, wikipedia and other data sources}
\label{SemanMathEnvSources}
In the last section examples of repositories related to materials where mentioned. There are of course a lot of other kinds of scientific repositories (see e.g. the list of open access repositories at opendoar.org\link{http://www.opendoar.org/}). These are however often not publiquely accessible, moreover if one takes texts (preprints, journals etc.) aside then the involved data formats are not always easy  accessible even not to a general scientific minded audience, that is explanations like

\vspace{2mm}
{\em Example of WYCKOFF-CAR (cut/paste it, then "wyckoff to poscar", then convert to standard primitive for the RHL cell, or standard conventional for the HEX cell): ...}

\vspace{2mm}
(to be found at the above mentioned aflowlib's front page \cite{aflowlib} \link{http://www.aflowlib.org/}) are understandable only to experts.

\vspace{2mm}
There are meanwhile frameworks which may assist in setting up scientific repositories, like e.g. the repositories support project \link{http://www.rsp.ac.uk/} (see e.g. their buddy scheme \link{http://www.rsp.ac.uk/help/buddy/}). However on their website the support project writes (31.8.12) \link{http://www.rsp.ac.uk/start/before-you-start/what-is-a-repository/}:

\vspace{2mm}
{\em However, some more complex objects (websites, advanced learning objects, 3D topographical representations and other data sets) do present a technological challenge.}

\vspace{2mm}

For getting a glimpse on the involved problems, lets at this point thus focus on a well-known publiquely accessible giant repository, namely at the crowd sourced data source Wikipedia. 

For example the infoboxes (i.e. the little boxes on the right side of a Wikipedia page, which contain for the case of a city the number of inhabitants etc.) \link{http://en.wikipedia.org/wiki/MOS:INFOBOX} contain already quite some structured information. Amongst others the above mentioned DBpedia project has tools to parse the wikipedia site, especially with regard to the structured content in the infoboxes and people can collaboratively assign ("map")  a RDF "vocabulary" (DBpedia) and graph description to the data description in the infobox, which can then be part of the DBpedia database (see mappings website \cite{mappings}\link{http://mappings.dbpedia.org/index.php/Main_Page}). So let's regard this a bit with respect to the application mentioned in the previous section. 

Since there is the description "density" of lets say iron in the infobox at wikipedia \link{http://en.wikipedia.org/wiki/Iron} (look under ``physical properties'') one could e.g. assign a DBpedia subject URI to "density", likewise one could assign an URI for "density of iron near room temperature" and a property URI for "has the $g \cdot cm^{-3}$ value". Furthermore one could think about e.g. assigning a triple ("density of iron near room temperature" , "has $g \cdot cm^{-3}$ value", 7.874). However one would eventually want to specify ``density of iron near room temperature'' in more detail. For the word "density" (and a few other items from the chembox (the wikipedia infobox for chemicals)) the mapping has been done, that is the word ``density'' has an URI in DBpedia. The datatypes are however at the time of writing not (yet) copied from the DBpedia Mappings Wiki to the DBpedia ontology.
Doing this for many materials would allow to gather data of materials (like densities) from DBpedia and visualize this data, like here plot the density of materials near room temperature in one diagram (given that one has an appropriate visualization tool for this task). So one could for example ``surf'' through DBpedia or wikipedia (eventually include some parsing and some prefiltering, especially for DBpedia) and store the corresponding semantic surfhistory as a semantic graph with a corresponding tool (see also the project pathway \cite{pathway} \link{http://pathway.screenager.be/}), select data within the graph which is concerned with the material data to be investigated, extract the wanted information (eventually with a facetted search) and visualize it with the correspondig mathematical tools. For the example this could mean that one "surfs to materials" then to for example "plastic" and "metal" and then searches for the densities of some plastics and metals and plots them together in a diagram. To some extend it is possible to surf the Yago database \cite{YagoNaga}, however also here sofar not much content with respect to math/physics/engineering has been added, moreover visualization issues like the investigation of the surfhistory are sofar only in a limited way possible. Note that this process may happen collaboratively, like that one engineer collects information about the density of a material, the other about carbon footprint etc. There are sofar however not so many collaborative tools, where such an approach could be realized. 
 
Likewise one could think about having or linking (in a standartized way) to datasets in Wikipedia, which contain e.g. a $1\times n$ dimensional array (i.e. an ordered list) of measured values of density over temperature.  Here one would already need some mathematical RDF  description for "$1\times n$ dimensional array", which probably actually exist in some ontology but probably, as said above, not necessarily in some explicitly mathematically informed RDF ontology. It should also be mentioned that since an "ordered list" is a rather basic graph it is currently discussed to be included in the serialization of RDF (like within JSON-LD). With the concept of {\em named graphs} \link{http://en.wikipedia.org/wiki/Named_graph} one could eventually also rather straightforward define something like "mathematical RDF {\em classes}'' rather in the sense of classes in computer programming. This issue will be adressed in the next subsection.

So alone with Wikipedia/DBpedia interesting things could be done. It is not clear to the author to what extend the inclusion of datasets like within the {\em datahub} \cite{datahub}\link{http://thedatahub.org/} is envisaged to be linked to wikipedia. There is a newer initiative called {\em wikidata} \cite{wikidata} \link{http://meta.wikimedia.org/wiki/Wikidata/Technical_proposal} which envisages to provide machine readable, structured data of Wikipedia and which envisages to present it in parallel to usual Wikipedia websites. According to the german computer magazine Heise Online \cite{wikidataHeise} \link{http://www.heise.de/newsticker/meldung/Wikidata-Wikipedia-bekommt-Faktendatenbank-1497264.html} Wikidata is sponsored with 1.3 million Euros by private sponsors, 650.000 Euros come from the Allen Institute for Artificial Intelligence, 325.000 Euros from the Gordon and Betty Moore Foundation and 325.000 Euros from the company Google. The project wikidata is currently fueled by a team of employed specialists and volunteers.

In the context of public-private initiatives it is - as a side remark - interesting to look at the countrywise volonteer participation within the project DBpedia like via the language mapping statistics of DBpedia at the DBpedia Mapping Sprint, Summer 2011 \cite{mappingssprint} \link{http://mappings.dbpedia.org/sprint/}. Here it seems on a first glance that Greece and Portugal are quite leading the race. The current economical situation in these countries may play a role here. 

It should also be mentioned that the author suspects that there seem to be efforts by the International Organization for Standartization to standartize semantic information (see ISO/IEC 19788, called MLR \cite{ISO} \link{http://isotc.iso.org/livelink/livelink/open/jtc1sc36}. The works are however not freely available. Parts can be purchased via ISO/IEC 19788-1:2011 \link{http://www.iso.org/iso/iso_catalogue/catalogue_tc/catalogue_detail.htm?csnumber=50772}. In their description on this site it is outlined:

\noindent {\em "The primary purpose of ISO/IEC 19788 is to specify metadata elements and their attributes for the description of learning resources. This includes the rules governing the identification of data elements and the specification of their attributes."}

There seems also to be a document directly concerned with automated content with the title: {\em N2448 - ``Summary of voting on ISOIEC NP 18343, Learning environment profile for automated contents.''} \link{http://isotc.iso.org/livelink/livelink/open/jtc1sc36}. The authors suspects are based on this ISO website informations, so the suspect could easily be wrong. 

\vspace{2mm}
As the author understood from the various discussions on the wikidata mailing list \link{http://news.gmane.org/gmane.org.wikimedia.wikidata} the costs of ISO documents are currently too high for an inclusion in the wikidata project. For commercial applications the ISO standards are rather important, so it is an important question in how far wikidata could relate to these standards. This may in particular involve legal issues.

Likewise (concerning the 3D modelling input) one could e.g. think of providing an RDF ontology for blender \cite{blender} \link{http://www.blender.org} in and output. As an example: Blender allows to set custom properties \link{http://wiki.blender.org/index.php/Doc:2.6/Manual/Extensions/Python/Properties} so for example if one could read in the chembox item "density" (like from DBpedia or when this is realized in wikidata) from lets say iron (for wikidata this could be done e.g. via an already existing python script called pywiki \link{https://github.com/jcreus/pywikidata}) then one could use this information for 
a custom property for computing the mass of a volume (as modelled
in Blender). This could then be read in e.g. as the mass of a 3D model into Blenders physics engine bullet \link{http://www.blender.org/development/release-logs/blender-240/bullet-physics/} but of course also for measuring the material use in an eco product design.

\section{RDF and automated reasoning}
\label{autoreason}
 In his article Massimo Marchiori \cite{Marchiori2003} \link{http://www.w3.org/People/Massimo/papers/2003/mathsemweb_mkm_03.pdf} wrote:

\vspace{2mm}
{\em ``And one of the possible cool functionalities that math functions an operators could have is just this: to be computable, in the sense that they could, somehow, be activated (evaluated), and the appropriate result returned.''}

\vspace{2mm}

He propsosed xquery \link{http://en.wikipedia.org/wiki/Xquery} as:

 \vspace{2mm}
{\em ``an example of a first catalog of functions and operators that are going to be standardized, for uniform usage over the Web. Similar efforts could be undertaken to extend such collection to more mathematical functions, where first-order functions and operators could be considered first. }

\vspace{2mm}
It is of course an interesting question in how far RDF descriptions of math could be made directly computable. Open math \cite{OpenMath}\link{http://en.wikipedia.org/wiki/OpenMath} can be seen as an approach to design a semantic description which is suited for tranlations to e.g. computer algebra systems. Apriori, as pointed out by Massimo Marchiori  \cite{Marchiori2003} \link{http://www.w3.org/People/Massimo/papers/2003/mathsemweb_mkm_03.pdf}, it seems that especially functional programming languages may be rather suitable for a translation, even for higher order functions. So it is to be asked in how far it could also be possible to include translations for compiler like infrastructures like LLVM \link{http://en.wikipedia.org/wiki/LLVM}, as these operate also on a rather functional level. It is to be expected that this could enable an even wider spread of the RDF descriptions.

The generality of RDF and in particular the possibility to link with an URI not only to a text description but in principle also to a whole RDF graph (this concept is called a {\em named graph} \link{http://en.wikipedia.org/wiki/Named_graph}) could allow not only for the description of higher level mathematical objects (like a function entails domain and range etc.) but eventually also for the translation to programming languages with a higher level of abstraction. A named graph could here be for example thought of as a kind of ``class'' like in object oriented programming (Interfaces, like in Java could probably also be modelled with a graph). The difference to a real programming language would however not only lie in the fact that the ``classes'' definitions could be ``spread over the internet'' but differences would also appear in the fact that the syntax itself (which in a program has to be known by the programmer) should apriori be described within RDF. So the semantic web could act in this case as a kind of IDE \link{http://en.wikipedia.org/wiki/Integrated_development_environment}. It is really an interesting question in how far this would allow for translations into languages like Java or Python. It should be noted at this place that a RDF graph related to higher level programming would - at least in the near term future -  rather serve for the description of processable entities than that it would form an executable code - alone for processing time and network traffic reasons. Apart from the technical limitations it is furthermore to be asked in how far one would really like to turn parts of the internet into executable code. The implications of such a big ``machine-brain'' should be discussed elsewhere.

Another useful feature of RDF math descriptions could however be the inclusion of complete code. Like for example in the above mentioned ab-initio calculations the corresponding mathematical source code (like for a Sage notebook) could be directly linked to via RDF. In particular RDF allows to identify the involved mathematical programming language and could incorporate the corresponding source code as text elements, which formatting can be described in RDF. A somewhat similar approach was the suggestion to link to an open math entity via RDF \link{http://www.w3.org/2001/sw/Europe/reports/xml_test_cases/wp53.html}. These approaches are of course way easier to implement than the other mentioned options for automated reasoning.

\section{RDF and visualizations, Mimirix and other examples}
\label{RDFviz}
There exist a couple of tools which allow to visualize RDF data also in non-tabular form, like in visual depictions of graphs.  Some of these allow also to collaboratively edit RDF data (or semantic web data in a different form) however as already mentioned above there are not overly many collaborative examples. It seems however that especially for engineering applications collaborations may be important. Within the production of a complex product, like a car there are usually many people involved.  The author of the original version of this essay was thus supervising a student project where such a tool has been created. The name of the tool is Mimirix, it is described in the following section. Other (collaborative) applications are described further down. The applications vary in their use of different computer languages and methodologies.

\subsection{Mimirix}
\label{mimirix}

\vspace{10mm}
\centerline{\includegraphics[height=.35\vsize,width=\hsize]{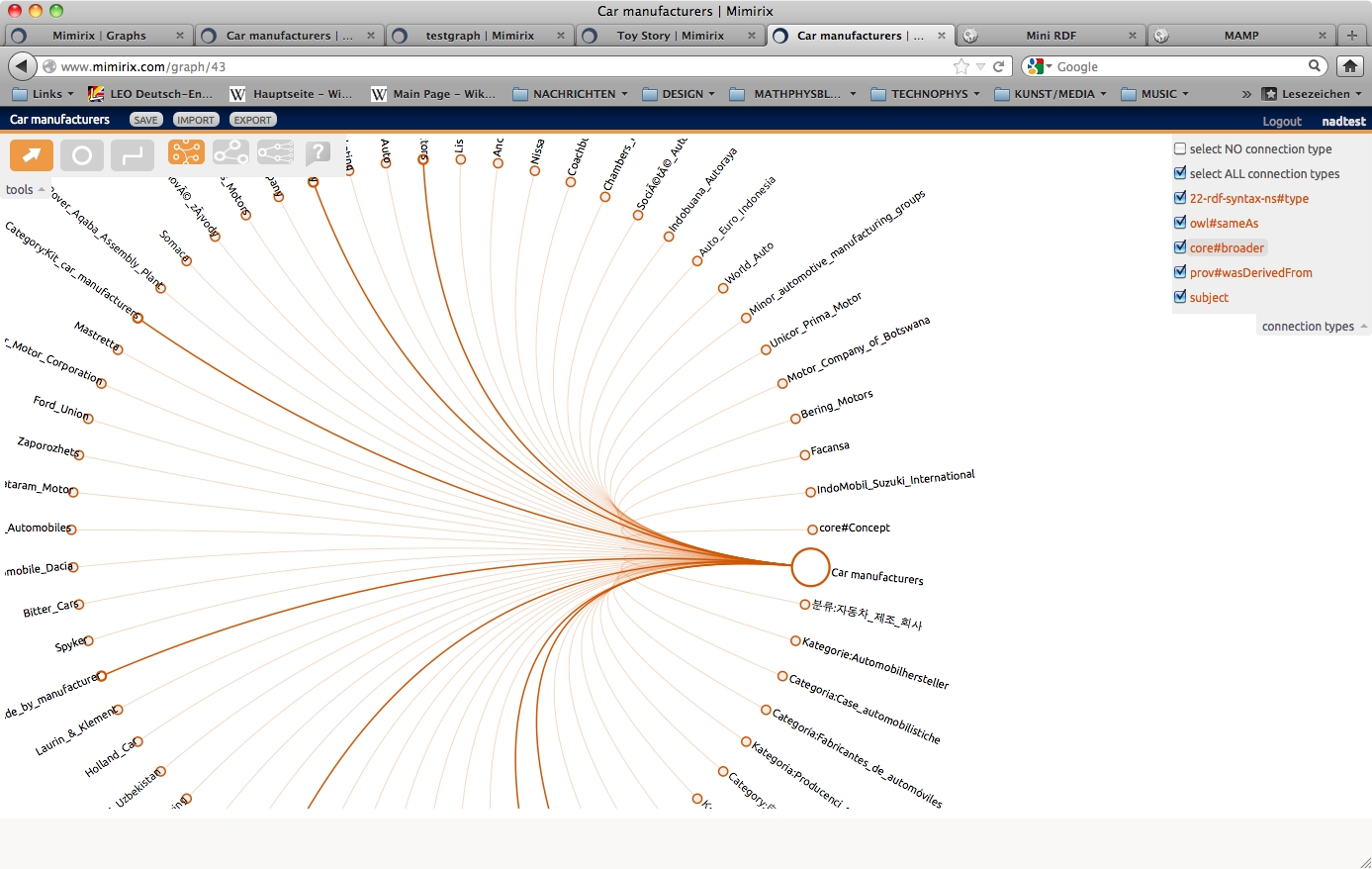}}
 \centerline{ \tiny screenshot of a DBpedia graph in "Circle mode" in Mimirix, the connection {\em core\#broader} is highlightened}

\vspace{10mm}
\centerline{\includegraphics[height=.35\vsize,width=\hsize]{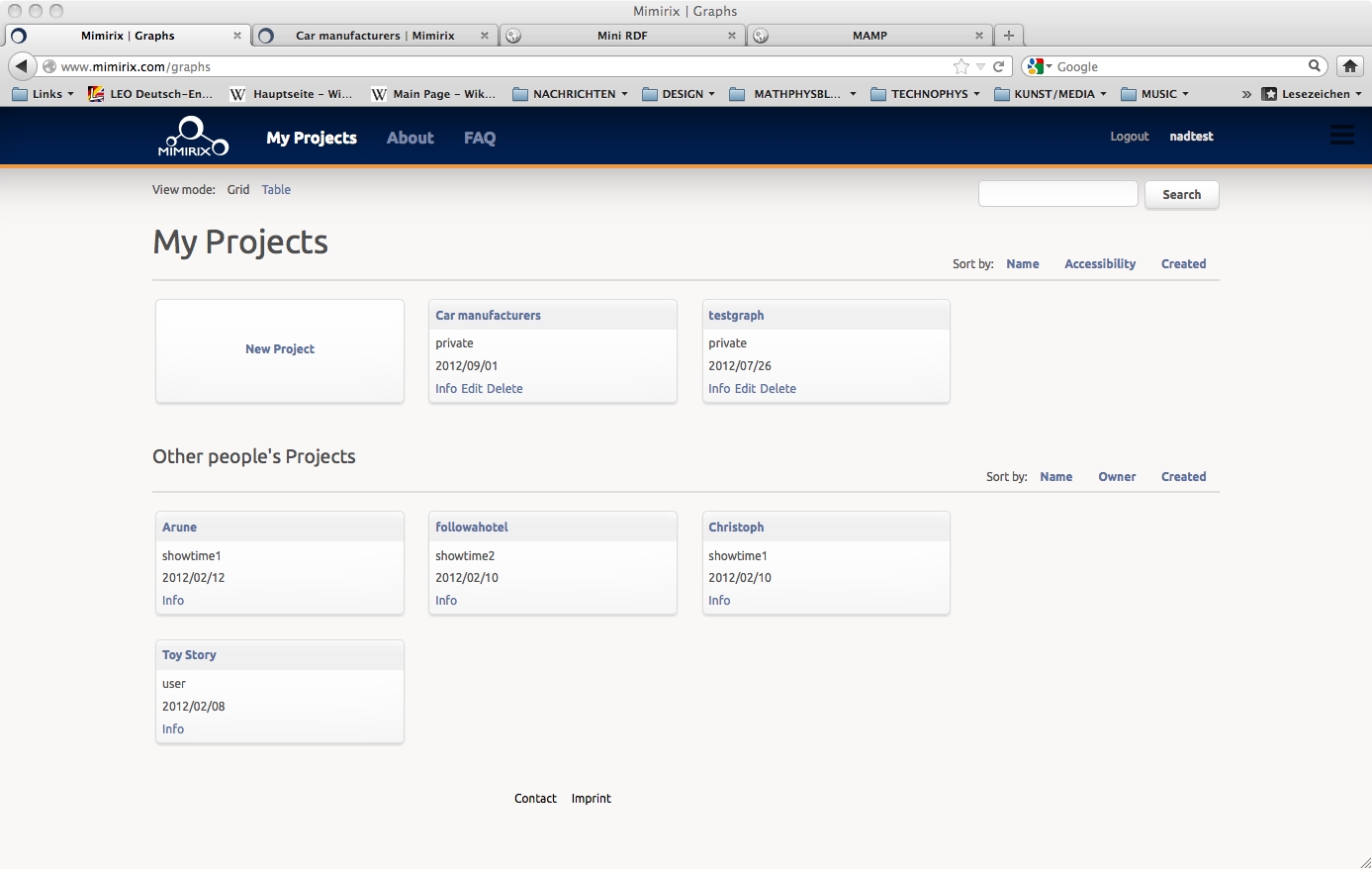}}
\centerline{\tiny screenshot of Mimirix interface for the user graph management in the socalled grid view} 

\vspace{10mm}
Mimirix \link{http://imi.htw-berlin.de/?page_id=2064} has been programmed as a Bachelor Project in the winter semester 2012 at the Berlin school Hochschule f\"ur Technik und Wirtschaft (HTW) by the students Thomas Alpers, Martin Bilsing, Igne Degutyte, Florian Demmer, Felix Griewald and Johanna von Rau\ss{}endorff. Mimirix is a collaborative environment which enables the processing of RDF data in a rather visual, intuitive and user friendly way. It is not adressing a semantic web informed audience, but suitable for a general audience. The frontend of Mimirix allows multiple users to log into the environment and to create and share RDF graphs.  A video by Florian Demmer which gives a demonstration about the functionality of the client can be found at the following URL \cite{FloVideo}\link{http://vimeo.com/36802055}. The software is envisaged to get some free/open source licence.

Given the circumstances (one semester bachelor project) the environment is of course still in a kind of experimental stage. In particular within the client application the filtering and the perceptional access facilities of RDF data are sofar only in a rather limited form available, like merging, pruning, zooming into and out of nodes and the distortion of subgraphs, or facetted searches etc. are not yet possible.  In particular it is for example also not yet possible to filter numerical RDF data and/or for example possible to visualize numerical data (possibly in connection to other data or data applications, like axis names, units, notepads, commentboxes, wikis etc.) within a client application. As pointed out above there is however anyways not much mathematically structured RDF data publicly available. Applications for the customization of the visualization of RDF data (like for example for representing nodes and links by photos, icons etc.) as well as an easier integration of custom clients are additional features, which can be likewise envisaged. This is in particular important as visualizations are amongst others rather dependend on individual aspects, like visibility, cultural and theme related specifications, graphic design awareness/literacy, taste etc.

The client of Mimirix is written in Javascript and uses amongst others the libraries d3 \cite{d3} \link{http://mbostock.github.com/d3/} and jQuery \cite{jQuery} \link{http://en.wikipedia.org/wiki/jQuery}, the frontend uses also jQuery and is hooked to a content management system (CMS) \link{http://en.wikipedia.org/wiki/Contact_management_system}, which had been taylored in CakePHP \cite{CakePHP} \link{http://en.wikipedia.org/wiki/CakePHP} to suit the special collaborative tasks of Mimirix. The backend consists of the CMS, a LAMP bundle \link{http://en.wikipedia.org/wiki/LAMP_(software_bundle)}, Debian \link{http://en.wikipedia.org/wiki/Debian} and an arc2 \cite{arc2} \link{https://github.com/semsol/arc2/wiki} triplestore (a database for RDF triples). An access control plugin for arc 2 was not (yet) finished.

The software is currently on a server of HTW. It is not publicly available, since the client is not yet safe enough against code injection and sofar no javascript expert could be found to help out with these very specific questions. There are other issues, like serverspace limitations, legal issues and economic considerations which haven't been yet sufficiently adressed in order to admit a full public access to Mimirix' collaborative environment. Mimirix can be installed on any local server and thus the collaborative environment is a aprori not bound to one special server and database. 

\subsection{DBpedia, YAGO, Freebase and others}
\label{dbpediaviz}
There are few (collaborative) interactive visualizations within DBpedia and YAGO (like the mentioned YAGO2 browser), however to the authors knowledge there are (apart from the mentioned browser) sofar no realizations of collaborative applications, which would also be suited for an general audience, which has less experience with semantic web techniques (winter 2011/12). 

There are some (interactive) visualizations for the above mentioned freebase \link{http://wiki.freebase.com/wiki/Visualizations} and there are eventually more underway. In particular freebase (some tools are proprietary) offers RDF access and a JSON based API. The freebase blog however has been inactive for more than a year, likewise the community website \link{http://wiki.freebase.com/wiki/Category:Community} holds currently an outdated warning sign. 

The above mentioned LOD2 project \cite{lod2} provides a comprehensive overview about its tools at their LOD2 technology stack \link{http://lod2.eu/WikiArticle/TechnologyStack.html}. Here for example semantic data conform browsers, wiki's, search tools etc. can be found. 

More tools can be found on semanticweb.org \cite{semanticweb} \link{http://semanticweb.org/}.  The semanticweb.org community seems to be linked to the Wikidata project. There is however, to the authors knowledge sofar (winter 2011/2012) no collaborative environment, which is comparable with Mimirix or the below mentioned Deepa Mehta project.

\subsection{Deepa Mehta and Mimirix}
\label{deepamehta}

The project Deepa Mehta \cite{DeepaMehta}\link{http://www.deepamehta.de} which is a 12 year old Berlin based non-profit organization has worked on visualizations of semantic data. It currently provides, like the DBpedia project or the Mimirix project, visualizations of semantic data which are hooked to a collaborative environment. People at Deepa Mehta currently think about using parts of Mimirix, since their underlying data model doesn't yet support RDF. The original motivation for the Deepa Mehta project was rather less motivated by the idea of having an easy accessible semantic web tool but rather motivated by in the idea of finding tools which are cognitively better acessible for coworking.  Deepa Mehta is GPL which could make the integration of further software development (like for the above mentioned material applications) into commercial applications difficult. This could thus be a handicap for cooperations with companies, which could for example provide more detailled data and services (like material or environmental data, like supply chain impacts (see e.g. the article on greenbiz ("Puma's Eco-Impacts Report Kicks the Ball Forward on Transparency \cite{greenbizPuma}\link{http://www.greenbiz.com/blog/2012/02/10/pumas-eco-impacts-kicks-ball-forward-transparency"} in exchange for the use of the software.  On the other hand if further software development is mainly done by volunteers as it had been the case for Deepa Mehta and to some extend also for Mimirix then it is of course to be asked why one should "donate" such a software to potential profit makers for free.

\section{Conclusion}

Information technology comes at a rather high price for humanhood and environment. Higher energy needs, the depletion of resources and the devaluation of human labour are only catch words which could indicate the scope of the involved problems.

This article was intended to give an outlook on possible applications in which information technology and in particular semantic web techniques may a priori be beneficial to humans and in particular to environmental efforts. How beneficial such an approach may be in the end depends of course in particular on the concrete implementations. 

\section{Acknowledgement}
A part of this work was done while being Lehrbeauftragter at HTW Berlin. I would like to thank some commentors at wikidata. Moreover I would like to thank my students at HTW for interesting discussions, I have also learned a lot from them, the people at Azimuth project for hosting pre-versions of this article and their feedback in general discussions and Tim Hoffmann for the financial and emotional support of this article.
\bibliographystyle{alpha}
\bibliography{semantic}
\end{document}